\documentstyle [12pt, epsf] {article}

\catcode`@=12 \topmargin 0.0in \evensidemargin 0.0in
\begin{document}
\vspace{0.5in}
\oddsidemargin -.1 in
\newcount\sectionnumber
\sectionnumber=0

\def\lsim{\mathrel{\vcenter{\hbox{$<$}\nointerlineskip\hbox{$\sim$}}}}
\thispagestyle{empty}
%%%%%%%%%%%%%%%%%%%%%%%%%%%%%%%%%%%%%%%%%%%%%%%%%%%%%%%%%%%%%%%%%%%%

\vskip.5truecm
\vspace*{0.5cm}

\begin{center}
{\Large \bf \centerline{Looking for forward backward asymmetries in
$B\to K \mu^+ \mu^-$} and $K^+\to \pi^+\mu^+\mu^-$}

\vskip3.5cm \centerline{\large{Chuan-Hung Chen}$^1$,
\large{Chao-Qiang Geng}$^2$ and \large{Anjan K. Giri}$^2$}
\centerline{$^1$ Physics Department, National Cheng Kung
University, Tainan, Taiwan 701, R.O.C} \centerline{$^2$ Physics
Department, National Tsing Hua University, Hsinchu, Taiwan 300,
R.O.C.}
\bigskip

\begin{abstract}

We investigate the possibility of scalar interaction affecting the
forward backward asymmetry in the decay mode $B\to K \mu^+ \mu^-$.
Using the scalar contribution and advocating Cheng-Sher type
ansatz we obtain sizable forward backward asymmetry. Furthermore,
we study the effect of the scalar interaction in $K^+ \to \pi^+
\mu^+ \mu^-$. It is pointed out that non-zero forward backward
asymmetry in $B\to K l^+ l^-$, if found in future B-factories,
may indicate the new physics in the scalar sector.

\end{abstract}
\end{center}

\thispagestyle{empty}
\newpage

Standard model is very successful, as verified by the data, but it
is still believed that physics beyond the standard model (SM)
might be around the corner. Experimental searches are on for the
same while we look into some avenue where we expect it to be
observable. In this context we would like to study the forward
backward asymmetries ($A_{FB}$) in $B\to K \mu^+ \mu^-$ and
$K^+\to\mu^+\mu^-$, with a hope that they might be observed in
future. The basic idea behind this study is that of the possible
new physics in the scalar sector, i.e., through the neutral Higgs
boson contribution leading to flavor changing neutral current
(FCNC) at the tree level. We are therefore interested to study the
forward backward (FB) asymmetries in the decay modes and some
related processes which are governed by the quark level processes
$d_i\to d_j l^+l^-\ (d_i=b,s,d)$. To begin with, we point out that
there exist many works in the literature involving these processes
in the standard model framework and also in many beyond the
standard model extensions [1-12]. So, we will not attempt to
repeat this here but mention the salient features and use basic
formulae needed, wherever necessary, to illustrate the idea.

It should be reminded here that forward backward asymmetries are
identically zero in the framework of the standard model for the
decay modes under consideration, i.e., in $B\to K\mu^+\mu^-$ and
$K^+\to\pi^+\mu^+\mu^-$, since scalar interactions are absent in
them and the forward backward asymmetry involves (see (\ref{fb}))
the scalar term. So in order to have nonzero forward backward
asymmetry we must have a scalar term, whose presence (that is
nonzero FB asymmetry in $B \to K \mu^+ \mu^-$) will give an
unambiguous signal of new physics in the scalar sector. We note
that in the most elegant extension of the standard model, i.e., in
MSSM one requires the extension of the Higgs sector and of course
the same in many extensions of the SM.

The standard model Higgs is the most sought after particle in the
upcoming collider experiments, which is the missing part in the SM
so far and is expected to be discovered very soon. Then Higgs
sector will be subjected to more stringent tests and possible
existence of the extension of Higgs sector may be confirmed
thereafter. On the other hand observation of nonzero FB asymmetry
in $B\to K \mu^+ \mu^-$ or $K^+\to\pi^+\mu^+\mu^-$ will provide an
invaluable clue to the existence of neutral Higgs boson
contribution in this decay mode and make inroads necessary for the
extension of the Higgs sector.

Unfortunately, we have not observed any $A_{FB}$ so far
\cite{babar1,belle1} but in the future B experiments FB asymmetry
will be investigated thoroughly and therefore we give another
close look at the possibility of tree level contribution to $b\to
s l^+ l^-$ process, keeping in mind the renewed interests
concerning the Z-induced FCNC in the literature and also the very
recent split supersymmetry idea, where one can have tree level
contributions. In order to implement the idea we use the idea of
Cheng and Sher for the FCNC. FCNC suppression is naturally taken
care of by the power form of the quark masses involved in the
process concerned (for details see \cite{cs}). It might be
possible that we have not been able to see this kind of
interaction because of the lower masses involves in the light
hadron sector but in the decays involving second and third
generation quarks, i.e., in $b\to s$ decays the masses being
larger so there is a possibility that it might be observable in
these decays, which is being subjected to tests in the currently
running B factories and so will continue in the future
B-experiments. Whether or not we will be able to see any existence
of tree level FCNC contribution will be verified in the future B
related experiments but at least from theoretical point of view we
would like to see whether it does make any sense or not, i.e.,
whether we can really have any indication of new physics, at least
with this kind of idea. We take the ansatz as $(m_i m_j)^{1/2}/v
$, where $m_{i,j}$ are the masses of the quarks involved and $v$
is the vev.

Here we present the relevant formulae used and provide some
details, which are necessary. The most general effective
lagrangian for the decay mode $B\to K \mu^+\mu^-$ can be written
as \cite{CQ}
\begin{equation}
{\cal M} = \frac{G_F \; \alpha}{\sqrt{2}\pi}\; V_{tb} V_{ts}^* \;[
F_S \bar l l +F_P \bar l \gamma_5 l +F_V p^\mu \bar l \gamma_\mu l
+F_A p^\mu \bar l \gamma_\mu \gamma_5 l ]\;,
\end{equation}
where $p^\mu$ is the four-momentum of the initial $B$ meson and
the $F_i$'s are functions of Lorentz-invariant quantities.
$F_{S,P}$ are absent in the SM but since our objective is to study
the effect of scalar interactions we deliberately keep them in our
calculation and would like to see the possible effects, if any. It
should be noted here that the tensor type interaction is not
independent since it can be reduced to a combination of scalar and
vector terms.

In this work our objective is to study the effect of scalar
interaction in the $A_{FB}$ in rare semileptonic B decays.
Therefore, we start with the effective lagrangian for $b\to s l^+
l^-$ including the new physics effect due to scalar type
interaction:
\begin{eqnarray}
{\cal H}_{eff}&=& -4 \frac{G_F}{\sqrt{2}} V_{tb} V_{ts}^* \left\{
\sum_{i=1}^{10} c_i (\mu) {\cal O}_i(\mu) +c_S (\mu) {\cal
O}_S(\mu) +c_P (\mu) {\cal O}_P(\mu)
\right.
\nonumber\\
&& ~~~~~~~~~~~~~~~~~\left.+c_S^\prime (\mu) {\cal
O}_S^\prime(\mu)+c_P^\prime (\mu) {\cal O}_P^\prime(\mu)
\right\}\;,
\end{eqnarray}
where, $c_i^{(\prime)} (\mu)$ and ${\cal O}_i^{(\prime)} (\mu)$
are the Wilson coefficients and local operators respectively. One
can recover the effective Hamiltonian of the standard model taking
the limit the new physics coefficients $c_{(S, P)}^{(\prime)} \to
0 $. The detailed expressions of the operators can be found in the
literature (see for example \cite{bobeth}).

We use the following hadronic matrix elements responsible for the
exclusive decay $B\to K l^+ l^-$ as
\begin{equation}
<K(k)|\bar s \gamma_\mu b|B(p)> = (2p-q)_\mu f_+(q^2) +
\frac{M_B^2 -M_K^2}{q^2}q_\mu[f_o(q^2) - f_+(q^2)]\;,
\end{equation}
\begin{equation}
<K(k)|\bar s i\sigma_{\mu \nu} q^\nu b|B(p)> = -[(2p-q)_\mu q^2
-({M_B^2 -M_K^2}q_\mu]\; \frac{f_T(q^2)}{M_B + M_K}\;,
\end{equation}
where, $q^\mu=(p-k)^\mu$ is the four-momentum transferred to the
dilepton system. Further, employing equation of motion for $s$ and
$b$ quarks, we obtain
\begin{equation}
<K(k)|\bar s b|B(p)> = \frac{M_B^2-M_K^2}{m_b-m_s} f_0(q^2)\;.
\end{equation}
The form factors defined above ($f_0$, $f_+$ and $f_T$) are
functions of the invariant mass of the dileptons. The two
dimensional spectrum is then given by
\begin{eqnarray}
\frac{1}{\Gamma_0} \frac{d\Gamma (B\to K l^+ l^-)}{ds\; d cos
\theta}&=&\lambda^{1/2}(M_B^2, M_K^2, s)\; \beta_l \{
s(\beta_l^2|F_S|^2 +|F_P|^2)\nonumber\\
&&+\frac{1}{4}\lambda (M_B^2, M_K^2, s) [ 1-\beta_l^2\
cos^2\theta](|F_A|^2+|F_V|^2) + 4m_l^2\;M_B^2|F_A|^2 \nonumber\\
&&+2 m_l[ \lambda^{1/2} (M_B^2, M_K^2, s) \beta_l Re(F_SF_V^*)cos
\theta
\nonumber\\
&& +(M_B^2-M_K^2 +s) Re(F_P F_A^*)] \}\,,
%\nonumber;,
\end{eqnarray}
where, $s=q^2=(p_{l^+}+p_{l^-})^2$. $\Gamma_0=\frac{G_F^2
\alpha^2}{2^9 \pi^5 M_B^3}|V_{tb} V_{ts}^*|^2$, $\lambda (a, b,
c)= a^2+b^2+c^2 -2(ab+bc+ac)$ and $\beta_l=\sqrt{1-4m_l^2/s}$.
Furthermore, we have  defined $\theta$ as the angle between three
momentum vectors $\vec p_{l^-}$ and $\vec p_s$ in the dilepton
center of mass system. Also note that the values $s$ and $\theta$
are bounded by $ 4m_l^2\le s \le (M_B^2 -M_K^2)^2$ and $ -1\le
cos\theta \le 1$, respectively .

The forward backward asymmetry \cite{ali} is defined as
\begin{equation}
A_{FB} (s)= \frac{ \int_0^1 d cos\theta \frac{d\Gamma}{ds d
cos\theta}- \int_{-1}^0 d cos\theta \frac{d\Gamma}{ds d
cos\theta}} { \int_0^1 d cos\theta \frac{d\Gamma}{ds d cos\theta}+
\int_{-1}^0 d cos\theta \frac{d\Gamma}{ds d cos\theta}}\;,
\end{equation}
which for the process under consideration is given by
\begin{equation}
A_{FB}(s)= \frac{2m_l \lambda (M_B^2, M_K^2, s) \beta_l^2
Re(F_SF_V^*) \Gamma_0}{d\Gamma/ds}\;.\label{fb}
\end{equation}
The dilepton invariant mass spectrum, $d\Gamma/ds$ can be obtained
by integrating the distribution (6) with respect to $cos \theta$,
which can be read as
\begin{eqnarray}
\frac{1}{\Gamma_0}\frac{d\Gamma (B\to K l^+ l^-)}{d s}&=& 2
\lambda^{1/2} (M_B^2, M_K^2, s) \beta_l \{s(\beta_l^2|F_S|^2 +|F_P|^2)\nonumber\\
&+&\frac{1}{6}\lambda (M_B^2, M_K^2, s) ( 1+2m_l^2/s)(|F_A|^2+|F_V|^2) \nonumber\\
&+& 4m_l^2\;M_B^2|F_A|^2  + 2 m_l(M_B^2-M_K^2 + s) Re(F_P
F_A^*)\}\;,
\end{eqnarray}
$F_i$'s used in the above formulae are combination of Wilson
coefficients and s-dependent functions ($f_i$'s), which are
\begin{eqnarray}
F_S & =&\frac{1}{2}(M_B^2- M_K^2) f_0(s) \left[ \frac{c_S m_b
+c_S^\prime m_s}{m_b-m_s}\right]\;,
\nonumber\\
%\end{equation}
%\begin{equation}
F_P&=& -m_l c_{10} \{ f_+(s) -\frac{M_B^2- M_K^2}{s}(f_0(s)
-f_+(s))\}
\nonumber\\
&&+ \frac{1}{2}(M_B^2-M_K^2) f_0(s)[\frac{c_P m_b +
c_P^\prime m_s}{m_b-m_s}]\;,
%\end{equation}
%\begin{equation}
\nonumber\\
F_A&=&c_{10} f_+(s)\,, ~~ F_V=[c_9^{eff} f_+(s) + 2 c_7^{eff} m_b
\frac{f_T(s)}{M_B + M_K}]\;.
\end{eqnarray}
In the above we have used the Wilson coefficients as
\begin{equation}
c_7^{eff} =-0.308, c_9=4.154,~ c_9^{eff} = c_9 + Y(s)+ C^{res},~
c_{10}=-4.261\;,
\end{equation}
 with
 \begin{eqnarray}
  Y(s) &=& g(m_c, s) (3c_1+c_2 +3c_3+c_4+3c_5+c_6
)-\frac{1}{2}g(m_s, s)(c_3+3c_4)
\nonumber\\
&&- \frac{1}{2}g(m_b, s)
(4c_3+4c_4+3c_5+c_6)+\frac{2}{9}(3c_3+c_4+3c_5+c_6)\,,
\nonumber\\
% The long distance effect included is $
C^{res}& =&
\frac{3\pi}{\alpha^2}(3c_1+c_2+3c_3+c_4+3c_5+c_6)\sum_{V_i =
\psi(1),\psi(2)}^{} \kappa \frac{\Gamma(V_i \to
l^+l^-)m_{V_i}}{m_{V_i}^2-s-im_{V_i}\Gamma_{V_i}}\,,
 \end{eqnarray}
where use has
been made of $ y_i = 4m_i^2/s $ and we have kept only two dominant
resonances. The functions $g(m_i, s)$ are as defined in
\cite{buras}.

The s-dependent form factors ($f_i$'s) used above are obtained
from the recent LCSR fit \cite{ball} as
\begin{eqnarray}
f_0(s)&=&\frac{0.1903}{1-s/39.38}\,,
 \nonumber\\
 f_+(s)&=&
\frac{0.3338}{1-s/29.3} +\frac{0.1478}{(1-s/29.3)^2}\,,
\nonumber\\
f_T(s)&=& \frac{0.1851}{1-s/29.3} +\frac{0.1905}{(1-s/29.3)^2}\,.
\end{eqnarray}

Using the formalism presented above, we would like to see the
possibility of observing nonzero FB asymmetry, and if found then
it will indicate clearly the presence of new physics in the scalar
sector. In order to visualize that we use the NNLO Wilson
coefficients and use the Cheng-Sher formalism to calculate the
appropriate scalar coefficient ($c_S$). Yukawa interaction between
scalar and fermion can naturally exist (see for example
\cite{davis}) and can be taken in the form
\begin{equation}
{\cal L} = \xi_{ij} ~\bar{Q}_{i, L}~ \phi_2 ~ D_{j, R} + (~... ~)+
h.c.\;,
\end{equation}
where the $...$ denotes the contribution from the up quark sector.
In the above the scalar doublet $\phi_2$ mediates the FCNC $d_i
\leftrightarrow d_j$ at the tree level, with nonzero $\xi_{ij}$.
In order to compute the FB asymmetry we have incorporated the idea
of Cheng and Sher \cite{cs} and define
\begin{equation}
\xi_{ij} = \lambda_{ij} (m_i m_j)^{1/2}/ v\;,
\end{equation}
where, $\xi$ is a dimensionless parameter, and $\lambda$'s are
expected to be order one. In fact the values of $\lambda$'s can be
taken in the range $\lambda_{bs} \sim 1-10$ and
$\lambda_{\mu\mu}\sim 1$. $B_s - \bar{B_s} $ mixing can provide
dominant contribution to the $\lambda_{bs}$ and thus it could be
bounded from that but with the available lower limit of $\Delta
M_{B_s} > 9.5 \times 10^{-12}$ GeV gives no constraint on
$\lambda_{bs}$, since the SM contribution exceeds this value. When
there will be a measurement then we will be able to constrain
$\lambda_{bs}$ more meaningfully. In the absence of that let us
try to explore other options at hand. In the literature $Z\to bs$
interaction vertex has been used to constrain the $\lambda_{bs}$
coupling and in Ref. \cite{atwood} it has been obtained that
$\lambda_{bs} < 10$, while for $\lambda_{\mu\mu}$ we will use the
value $\lambda_{\mu\mu} \simeq 1$. On the other hand in Ref.
\cite{bobeth} using model independent analysis it has been
inferred (using the upper limit of $B\to K^* \mu^+\mu^-$ ) that
the scalar/pseudoscalar coefficient can be at most $R_{S,P} =4$
(in fact the set of $R_S=4$ (maximum) and $R_P=0$ was obtained to
be the most preferable option), where $R_S \equiv c_i/c_i^{SM}$,
and $R_{S,P} \equiv m_b c_{S,P}$. $c_i^{SM}$ are the standard
model Wilson coefficients and $c_{S,P}$'s are as defined before.
Recently, a new upper limit has been provided by Tevatron/CDF for
the $B_s \to \mu^+\mu^-$ ($B_s \to \mu^+\mu^- < 5.8 \times
10^{-7}$ \cite{cdf}) and we use the same to constrain
$\lambda_{bs}$, which is found to be $\lambda_{bs} \sim $ 3.5.
Furthermore, we recheck that this is well within the maximum limit
obtained in \cite{bobeth}. In this analysis we have used the value
of the scalar Higgs mass to be 150 GeV.

It should be noted that in [6] the authors have studied the effect
of Higgs boson contribution on the $A_{FB}$ in two Higgs doublet
model (THDM)-type-II and MSSM with minimal flavor violation and
large ${\rm tan} \beta$, with $ 40\le {\rm tan} \beta \le 60$.
Moreover, the ${\rm tan} \beta $ dependence is strong which
appears as ${\rm tan}^2 \beta$ and the effect of the neutral Higgs
boson is taken at the loop level. In contrast, here we have
considered the tree level scalar contribution ($\xi_{bs}
\xi_{\mu\mu}/ m_H^2$) for the $A_{FB}$ with the Cheng-Sher type
ansatz. Our calculation is based on tree level contribution
instead of loop effects as in [6] and also we have not used any
strong dependence like ${\rm tan}^2 \beta$ (with $40 \le {\rm tan}
\beta \le 60$). Although we have not considered any specific model
but the tree level contribution is similar to THDM (type-III). In
order to constrain the FCNC coupling we have used the latest
result on $B_s \to \mu^+\mu^-$ and also checked that it does not
contradict any existing bound. Before proceeding further, we would
like mention here that if nonzero $A_{FB}$ is found then it may
not be immediately possible to infer whether it corresponds to
tree level contribution or a loop effect and hence more careful
studies are required to distinguish the same. But given the
simplicity of our model and no strong dependence on any model
dependent parameter (apart from the FCNC, which is ensured to be
very small with the use of Cheng-Sher type ansatz) our explanation
will be preferred over the other explanations. In a sense it may
be meaningful to say that it may indicate to the one like
THDM-(type-III), but not necessarily.

Using the parameters mentioned above and using the relevant input
parameters (like, $m_B$=5.28, $m_K$=0.5, $m_\pi$=0.14, $m_b$=4.6,
$m_s$=0.15, $m_\tau$=1.78 and $m_\mu$=0.105 in units of GeV
\cite{pdg}, $\alpha$=1/128) we plot the forward backward asymmetry
as a function of the invariant mass of the dileptons, which is
presented below. The branching ratio for $B\to K \mu^+\mu^-$ is
obtained to be $6.8 \times 10^{-7} $, including scalar interaction
(as compared to the experimental value $5.6^{+2.9}_{-2.4} \times
10^{-7}$ \cite{pdg}).

\begin{figure}[htb]
   \centerline{\epsfysize 2.0 truein \epsfbox{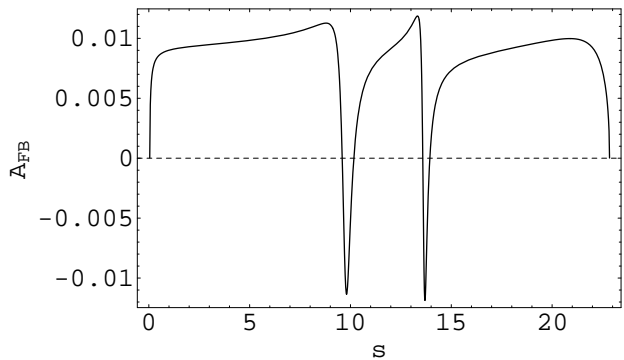}}
 \caption{
  The  forward backward asymmetry ($\rm{A_{FB}}$) in $B\to K \mu^+\mu^-$,
  where $\rm{s}$ is in $\rm{GeV}^2$ }
  \end{figure}

From the above figure, it becomes evident that we can expect to
have some nonzero FB asymmetry in $B\to K \mu^+\mu^-$, which is a
consequence of the fact that there might be tree level scalar
interactions present in $b\to s $ transition. It can be seen from
the figure that $A_{FB}$ can arise at the level of few percent due
to the scalar type interaction and the integrated FB asymmetry
over the whole dilepton invariant mass is found to be around a
percent level. We have also studied the affect of scalar type of
interactions in the forward backward asymmetry in $B\to K \tau^+
\tau^-$, which is shown below. It can be seen that here too one
can have sizable FB asymmetry and it can be tested in future. The
branching ratio obtained for $B\to K \tau^+\tau^-$ is $30 \times
10^{-7}$.
\begin{figure}[htb]
   \centerline{\epsfysize 2.0 truein \epsfbox{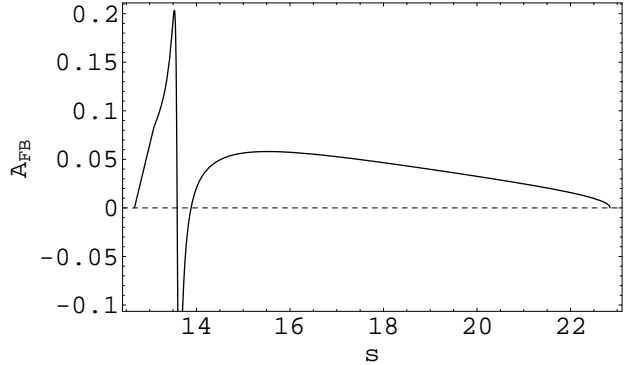}}
 \caption{
  The  forward backward asymmetry ($\rm{A_{FB}}$) in $B\to K \tau^+\tau^-$,
  where $\rm{s}$ is in $\rm{GeV}^2$}
  \end{figure}
Finally, we note that there are no significant deviations for the
FB asymmetries in $B\to K^*\mu^+\mu^-$ and  $B\to X_s l^+l^-$ from
that of the SM expectations.

With the above findings keeping in mind in the $b\to s $
transitions we thereafter looked into another similar possibility
in $s \to d$ type transition. Here, we studied to see the effect
of this type of scalar interaction in $K^+\to\pi^+\mu^+\mu^-$.
Since this process is also well studied in the literature, we do
not mention here the details explicitly. Following \cite{chen1}
and using the same idea, as above, we plotted the forward backward
asymmetry in the decay $K^+\to \pi^+ \mu^+\mu^-$ as a function of
the lepton invariant mass. Using scalar interactions and
Cheng-Sher ansatz we found ($f_S \sim 1\times 10^{-6}$) and used
$f_V = a_+ + b_+ \frac{s}{m_K^2}+ \omega (s)$, where $a_+$, $b_+$
and $\omega $ are parameters as used in \cite{chen1}, which are
extracted from the experimental data. Here we have used the value
of $\lambda_{ds} \sim $ 0.08, which is extracted from the
$K^0-\bar{K}^0$ mixing and $\lambda_{\mu\mu} \sim 1$. From the
figure it can be seen that nonzero $A_{FB}$ can be obtained with
scalar type interaction. It should be noted here that $A_{FB}
(K^+\to \pi^+\mu^+\mu^-) = O (10^{-3}) $ is accessible to future
experiments such as the CKM experiment at Fermilab \cite{hyper}.
The branching ratio obtained for $K^+\to \pi^+\mu^+\mu^-$ is $ 8.3
\times 10^{-8} $ (whereas the experimental value is $(8.1 \pm 1.4)
\times 10^{-8}$ \cite{pdg}) .

\begin{figure}[htb]
   \centerline{\epsfysize 2.0 truein \epsfbox{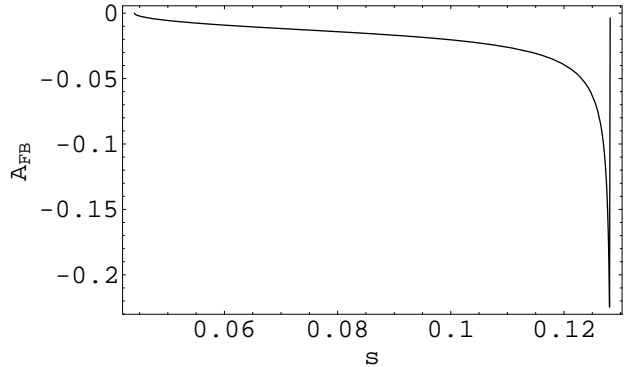}}
 \caption{
  The  forward backward asymmetry ($\rm{A_{FB}}$) in $K^+\to \pi^+
  \mu^+\mu^-$, where $\rm{s}$ is in $\rm{GeV}^2$ }
  \end{figure}

To summarize, in this brief note we have reanalyzed the
possibility of observing the new physics with scalar type
interactions in B decays. Forward backward asymmetry in $B\to K
\mu^+\mu^-$ is identically zero in the standard model. In order to
have non-zero FB asymmetry in $B\to K \mu^+\mu^-$ one must have
scalar type of interaction, which is absent in standard model
framework. Using Cheng-Sher type ansatz we have shown that one can
obtain non-zero FB asymmetry and thus if observed then it may
establish the presence of new physics, which unfortunately we have
not been able to observe so far. We obtain non-zero $A_{FB}$ in
$B\to K \mu^+\mu^-$ and in fact it is very significant in $B \to K
\tau^+\tau^-$.

The $A_{FB}$ in $K^+\to \pi^+\mu^+\mu^-$ like in the case of $B\to
K \mu^+ \mu^-$ and $B\to K \tau^+\tau^-$ seems to be also sizable
and could be detectable. In fact, the non-zero $A_{FB}$ in $B\to K
\tau^+\tau^-$ will be one of the best tools, in near future, to
establish the new physics. It should be pointed out that in order
to detect 2\% forward backward asymmetry, say in $B\to K
\mu^+\mu^-$ with branching ratio at the level of $\sim 6 \times
10^{-7}$ and at 3$\sigma$ level, around $10^{10}$ B's are needed.
To conclude, we note the fact that forward backward asymmetry in
$B\to K l^+l^-$ ($l=\mu, \tau $) and $K^+\to \pi^+\mu^+\mu^-$ is a
powerful tool to observe new physics. We hope, in the coming
years, it will be possible to observe the non-zero $A_{FB}$ and in
that case it may establish the existence of scalar interaction,
otherwise, the scalar interaction will be severely constrained as
a source of new physics, at least for the rare B and K decays.

This work has been supported by the National Science Council of
the Republic of China under Contract Nos:
NSC-93-2112-M-006-010,
NSC-93-2112-M-007-014 and
NSC-93-2811-M-007-060.

%%%%%%%%%%%%%%%%%REFERENCES%%%%%%%%%%%%%%%%%%%%%%%%%%%

\end{document}